\documentclass[twocolumn]{aastex6}

\usepackage{enumitem}

\hypersetup{citecolor=blue,linkcolor=blue,urlcolor=blue}
\usepackage[hang]{footmisc}

\renewcommand{\textsc}{}

\begin{document}

\title{Early Blue Excess from the Type I\lowercase{a} Supernova 2017\lowercase{cbv}\\ and Implications for Its Progenitor}

\author{
Griffin~Hosseinzadeh\altaffilmark{1,2},
David~J.~Sand\altaffilmark{3,4},
Stefano~Valenti\altaffilmark{5},
Peter~Brown\altaffilmark{6},
D.~Andrew~Howell\altaffilmark{1,2},
Curtis~McCully\altaffilmark{1,2},
Daniel~Kasen\altaffilmark{7,8},
Iair~Arcavi\altaffilmark{1,2,11},
K.~Azalee~Bostroem\altaffilmark{5},
Leonardo~Tartaglia\altaffilmark{3,4,5},
Eric~Y.~Hsiao\altaffilmark{9},
Scott~Davis\altaffilmark{9},
Melissa~Shahbandeh\altaffilmark{9}, and
Maximilian~D.~Stritzinger\altaffilmark{10}
}
\affil{
\altaffilmark{1}{Las Cumbres Observatory, 6740 Cortona Drive, Suite 102, Goleta, CA 93117-5575, USA; \href{mailto:griffin@lco.global}{griffin@lco.global}}\\
\altaffilmark{2}{Department of Physics, University of California, Santa Barbara, CA 93106-9530, USA}\\
\altaffilmark{3}{Department of Astronomy/Steward Observatory, 933 North Cherry Avenue, Rm. N204, Tucson, AZ 85721-0065, USA}\\
\altaffilmark{4}{Department of Physics \& Astronomy, Texas Tech University, Box 41051, Lubbock, TX 79109-1051, USA}\\
\altaffilmark{5}{Department of Physics, University of California, 1 Shields Avenue, Davis, CA 95616-5270, USA}\\
\altaffilmark{6}{Mitchell Institute for Fundamental Physics and Astronomy, Texas A\&M University, College Station, TX 77843-4242, USA}\\
\altaffilmark{7}{Nuclear Science Division, Lawrence Berkeley National Laboratory, Berkeley, CA 94720-8169, USA}\\
\altaffilmark{8}{Departments of Physics and Astronomy, University of California, Berkeley, CA 94720-7300, USA}\\
\altaffilmark{9}{Department of Physics, Florida State University, 77 Chieftain Way, Tallahassee, FL 32306-4350, USA}\\
\altaffilmark{10}{Department of Physics and Astronomy, Aarhus University, Ny Munkegade 120, DK-8000 Aarhus C, Denmark}
}
\received{2017 June 27}
\revised{2017 July 31}
\accepted{2017 August 2}
\published{2017 August 14}
\slugcomment{{\sc The Astrophysical Journal Letters}, 845:L11 (8pp), 2017 August 20}
\shorttitle{Early Blue Excess from the Type Ia SN 2017cbv}
\shortauthors{Hosseinzadeh et al.}

\begin{abstract}
We present very early, high-cadence photometric observations of the nearby Type~Ia SN~2017cbv.  The light curve is unique in that it has a blue bump during the first five days of observations in the $U$, $B$, and $g$ bands, which is clearly resolved given our photometric cadence of 5.7~hr during that time span.  We model the light curve as the combination of early shocking of the supernova ejecta against a nondegenerate companion star plus a standard SN~Ia component. Our best-fit model suggests the presence of a subgiant star $56\,R_\sun$ from the exploding white dwarf, although this number is highly model-dependent.  While this model matches the optical light curve well, it overpredicts the observed flux in the ultraviolet bands. This may indicate that the shock is not a blackbody, perhaps because of line blanketing in the UV. Alternatively, it could point to another physical explanation for the optical blue bump, such as interaction with circumstellar material or an unusual nickel distribution.  Early optical spectra of SN~2017cbv show strong carbon (\ion{C}{2} $\lambda$6580) absorption up through day $-13$ with respect to maximum light, suggesting that the progenitor system contains a significant amount of unburned material. These early results on SN~2017cbv illustrate the power of early discovery and intense follow-up of nearby supernovae to resolve standing questions about the progenitor systems and explosion mechanisms of SNe~Ia.
\end{abstract}

\section{Introduction}
\setcounter{footnote}{10}
\footnotetext{Einstein Fellow}

Type Ia supernovae (SNe~Ia) are the thermonuclear explosions of carbon--oxygen white dwarfs and are standardizable candles vital for cosmological distance measurements. Despite intense study, the progenitor scenarios and explosion mechanisms for these events are still not understood, and may have multiple pathways \citep[see][for reviews]{Howell11,Maoz14}.  The two main progenitor pictures are the single-degenerate (SD) scenario, where the white dwarf accretes material from a nondegenerate secondary star \citep{Whelan73}, and the double-degenerate scenario (DD), where two white dwarfs are present in the pre-supernova system \citep{Iben84,Webbink84}. However, the details of both scenarios are still under investigation \citep[e.g.,][]{Pakmor12,Kushnir13,Shen14,Levanon17}.

The early light curves of SNe~Ia are promising ways to constrain the progenitor systems and the physics of the explosion.  For instance, the collision of the SN ejecta with a nondegenerate companion star may manifest as an early blue or ultraviolet (UV) bump in the light curve, depending on the viewing angle \citep{Kasen10}. Early temperature or luminosity measurements can directly constrain the radius of the progenitor \citep{Piro10,Rabinak12}; observed limits have confirmed that the exploding star must be a white dwarf \citep[e.g.,][]{Nugent11,Bloom12,Zheng13}. Recently, \citet{Piro16} explored how early SN~Ia light curve behavior depends on the amount and extent of circumstellar material (CSM) and the distribution of $^{56}$Ni in the ejecta, which is expected to vary considerably with the location(s) of ignition in the progenitor.  Finally, different explosion mechanisms may also produce distinct early light curves \citep{Noebauer17}.

Very early SN~Ia light curve observations are becoming more common, and recent studies have shown that they display a range of early behaviors \citep{Hayden_etal_2010_shock,Bianco11,Mo_etal_2011,Brown_etal_2012_shock,Zheng13,Zheng14,Cao15,Firth15,Goobar15,Im15,Marion16,Zheng16,Shappee16}.  iPTF14atg \citep{Cao15} and SN~2012cg \citep{Marion16} both showed early, UV/blue excesses in their light curves.  iPTF14atg was an SN~2002es-like event, which are subluminous, do not follow the \cite{Phillips93} relation, have low velocities, and show \ion{Ti}{2} absorption \citep{Ganeshalingam12}. \cite{Cao16} found that the UV excess in iPTF14atg was consistent with the SN ejecta interacting with a companion star, although \cite{Kromer16} find the SD scenario incompatible with the observed spectral evolution.  In SN~2012cg, a normal SN~Ia, the early blue bump was again interpreted as a signature of ejecta--companion interaction, consistent with a $6\,M_\sun$ main-sequence companion $29\,R_\sun$ from the white dwarf, using the \citet{Kasen10} formulation.\footnote{Note that the models of \cite{Kasen10} directly constrain the binary separation, not the companion mass. Masses can be inferred by assuming the companion is in Roche lobe overflow and applying a mass-radius relationship.} However, \cite{Shappee16b} find that other probes of the progenitor system do not support the SD interpretation for SN~2012cg.

Here, we present the early light curve and spectra of SN~2017cbv, which show a clear blue excess during the first several days of observations. This excess may be a more subtle version of that seen in SN~2012cg and iPTF14atg, but seen more clearly here thanks to denser sampling.

\section{Observations and Data Reduction}
SN~2017cbv (a.k.a.\ DLT17u) was discovered on MJD 57822.14 (2017 March 10 UT) at a magnitude of $R \approx 16$ by the Distance Less Than 40~Mpc survey (DLT40; L.~Tartaglia et al.\ 2017, in preparation), a one-day cadence SN search using a PROMPT 0.4~m telescope \citep{prompt}, and was confirmed by a second DLT40 image during the same night \citep{2017TNSTR.294....1V}. Our last nondetection was on MJD 57791. The SN is located on the outskirts of the nearby spiral galaxy NGC~5643. Within hours of discovery (MJD 57822.7), the transient was classified as a very young SN~Ia with the robotic FLOYDS spectrograph mounted on the Las Cumbres Observatory \citep[LCO;][]{Brown_2013} 2~m telescope in Siding Spring, Australia \citep{2017TNSCR.300....1H}.

\defcitealias{sdssdr13}{SDSS Collaboration (2016)}
\textit{UBVgri} follow-up observations were obtained with Sinistro cameras on LCO's network of 1~m telescopes. Using \texttt{lcogtsnpipe} \citep{Valenti_2016}, a PyRAF-based photometric reduction pipeline, we measured aperture photometry of the SN. Because the SN is bright and far from its host galaxy, image subtraction and PSF fitting are not required. Local sequence stars were calibrated to the L101 standard field observed on the same night at the same observatory site using \textit{UBV} Vega magnitudes from \cite{Stetson00} and \textit{gri} AB magnitudes from the \citetalias{sdssdr13}.

The \textit{Swift} satellite began observing SN~2017cbv on MJD 57822.52.  Ultra-Violet Optical Telescope \citep[UVOT;][]{Roming_2005} photometry is given in the UVOT Vega photometry system using the pipeline for the \textit{Swift} Optical Ultraviolet Supernova Archive \citep[SOUSA;][]{Brown_etal_2014_SOUSA} and the zeropoints of \citet{Breeveld10}.  Smaller hardware windows were used in the optical near maximum brightness in order to reduce coincidence loss and measure brighter magnitudes without saturation \citep{Poole_etal_2008,Brown_etal_2012_11fe}.  A few $U$- and $B$-band observations on the rising branch were saturated and are excluded.

Finally, we obtained absolute magnitudes by applying the distance modulus, $\mu = 31.14 \pm 0.40$~mag ($16.9 \pm 3.1$~Mpc), of \citet{1988ngc..book.....T} and the Milky Way extinction corrections, $E(B-V) = 0.15$~mag, of \citet{Schlafly_2011}. We assume no additional host galaxy extinction, given the SN's position in the outskirts of NGC~5643 and a lack of narrow \ion{Na}{1}~D absorption in high-resolution spectra (D.J. Sand et al.\ 2017, in preparation).

\begin{figure*}
\includegraphics[width=\textwidth]{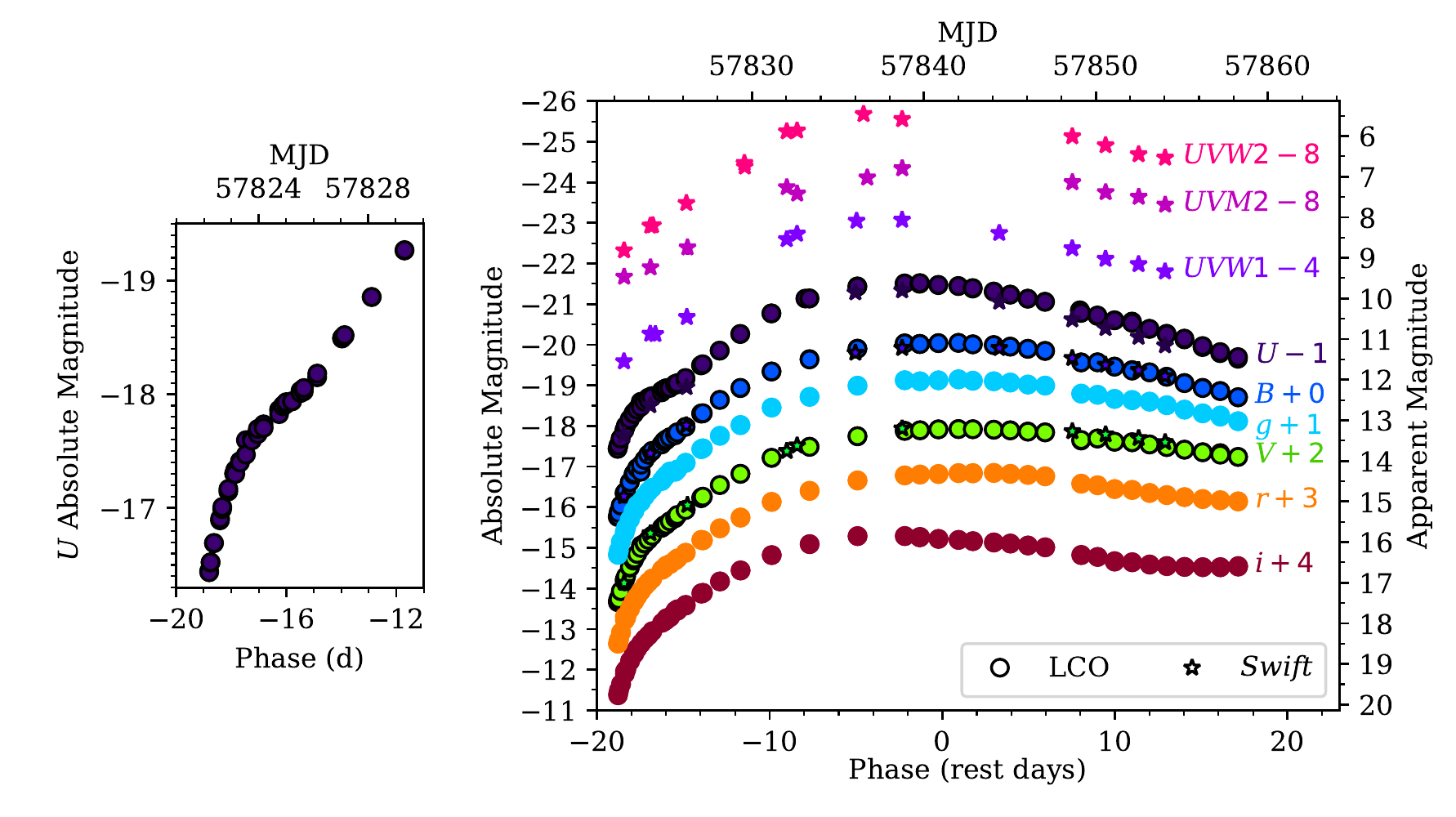}
\caption{\scriptsize UV and optical photometry of SN 2017cbv, in absolute and extinction-corrected apparent magnitudes. The left panel shows a bump in the early $U$-band light curve. (The data for this figure are available.)\label{fig:phot}}
\end{figure*}

Our photometry is shown in Figure~\ref{fig:phot}. By fitting quadratic polynomials to the observed light curve, one around peak and one around +15~days, we find that SN~2017cbv reached a peak magnitude of $B=11.72$~mag ($M_B=-20.04$~mag) on MJD 57841.07, with $\Delta m_{15}(B)=1.06$~mag. The decline rate of SN~2017cbv is near the average for normal SNe~Ia \citep[see, e.g., Figure~14 of ][]{Parrent14}. The peak absolute magnitude appears to be on the bright end of SNe~Ia \citep{Parrent14}, but this is uncertain due to the poorly constrained distance to NGC~5643. All figures use MJD 57821 as the nominal explosion date (see \S\ref{sec:spec}).

\begin{figure*}
\centering
\includegraphics[height=0.75\textheight]{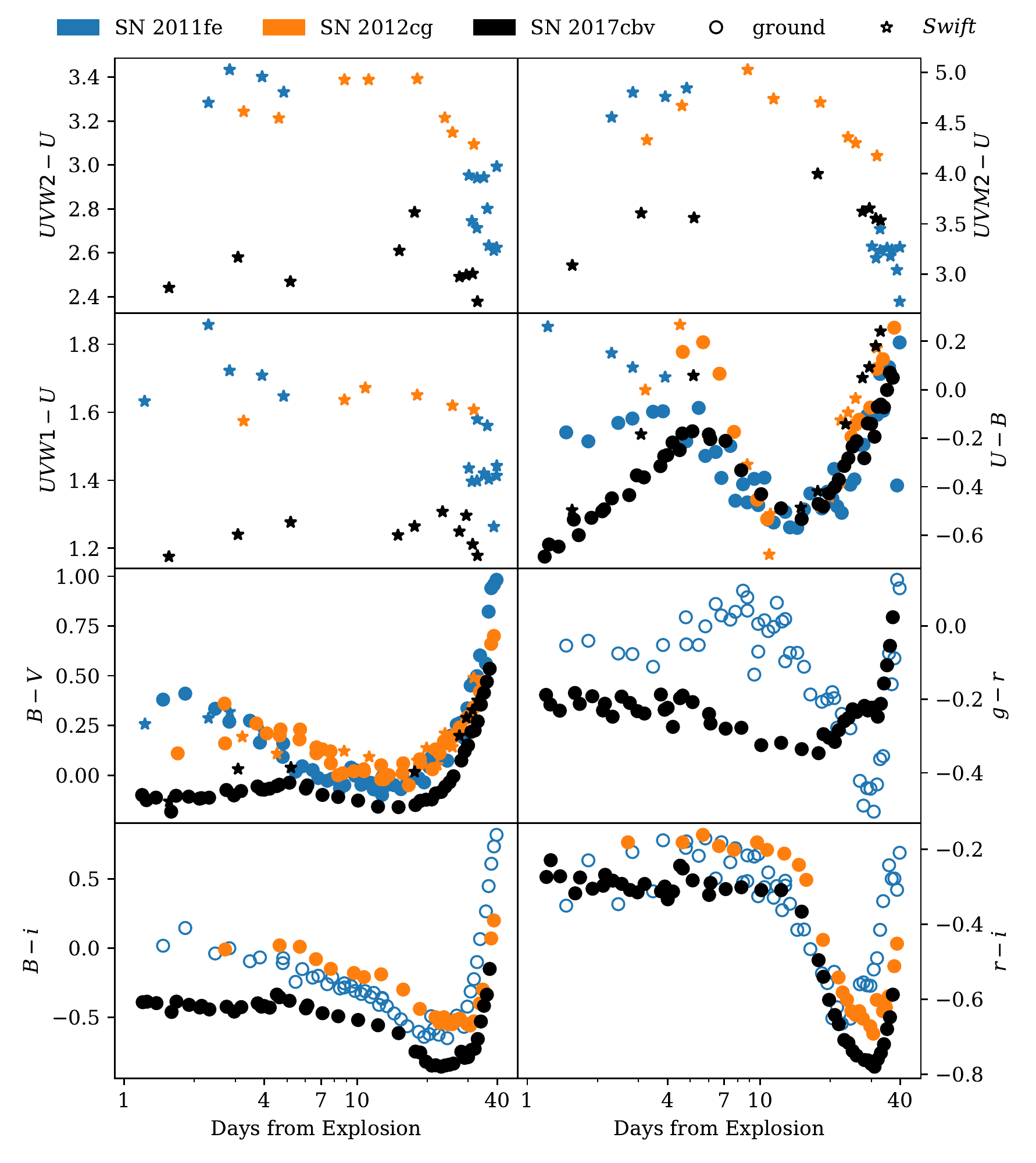}
\caption{\scriptsize Milky Way extinction-corrected ultraviolet and optical colors of SN~2017cbv, compared to SN~2011fe \citep{Zhang16} and SN~2012cg \citep{Marion16}. The colors of the subluminous SN~Ia iPTF14atg are quite different than colors of the other events \citep{Cao15}, so they are not shown. All \textit{Swift} photometry is from SOUSA \citep{Brown_etal_2014_SOUSA}. Open circles indicate $V-R$, $B-I$, and $R-I$ colors that have been converted to $g-r$, $B-i$, and $r-i$, respectively, using the transformations of \cite{Jordi2006}. Note the unusually blue $U-B$ and $B-V$ colors of SN~2017cbv during the bump (one to five days after explosion).\label{fig:colors}}
\end{figure*}

\section{Light Curve Morphology and Fitting}\label{sec:lc}
The very early light curve of SN~2017cbv shows a prominent blue bump in the $U$, $B$ and $g$ bands during the first five days of observation, also visible in its $U-B$ and $B-V$ colors (Figure~\ref{fig:colors}). This indicates a high-temperature component of the early light curve in addition to the normal SN~Ia behavior.

There are several possibilities for the origin of the early light curve bump (see \S\ref{sec:discuss} for a discussion), but here we fit the analytic models of \citet{Kasen10} for a nondegenerate binary companion shocking the SN ejecta, as might be expected from the SD scenario. While in reality such a collision would be highly asymmetric, we use \citeauthor{Kasen10}'s analytic expressions for the luminosity within an optimal viewing angle. By further assuming that the shock consists of a spherical blackbody, \citeauthor{Kasen10} arrives at the following equations for the photospheric radius and effective temperature\footnote{Equation~(\ref{eq:Teff}) corrects the exponent on $\kappa$ in \citeauthor{Kasen10}'s Equation~(25).}:
\begin{equation} R_\mathrm{phot} = (2700\,R_\sun) x^{1/9} \kappa^{1/9} t^{7/9} \label{eq:Rphot}\end{equation}
\begin{equation} T_\mathrm{eff} = (25{,}000\,\mathrm{K}) a^{1/4} x^{1/144} \kappa^{-35/144} t^{-37/72} \label{eq:Teff}\end{equation}
where $a$ is the binary separation in units of $10^{11}$~m ($144\,R_\sun$), $\kappa$ is the opacity in units of the electron scattering opacity (we fix $\kappa=1$ in our fits), and $t$ is the time since explosion in days.  In addition, we define
\begin{equation} x \equiv \frac{M}{M_\mathrm{Ch}} \left(\frac{v}{10{,}000\,\mathrm{km\,s^{-1}}}\right)^7, \label{eq:x}\end{equation}
where $M$ is the ejected mass, $M_\mathrm{Ch} = 1.4\,M_\sun$ is the Chandrasekhar mass, and $v$ is the transition velocity between power laws in the density profile.

Our light curve model is the sum of the SiFTO template \citep{sifto} to account for the normal SN~Ia emission, and \citeauthor{Kasen10}'s shock model to account for the blue excess. We scale each band of the SiFTO template independently. In the $U$, $B$, $V$, and $g$ bands, we fix the scaling factor to match the observed peak. In the $r$ and $i$ bands, we leave the scaling factor as a free parameter in order to account for any contribution from the shock in those bands around peak (predicted for some combinations of parameters). We also allow the time of $B$-band maximum light and the stretch \citep{Perlmutter97} of the SiFTO template to vary.

In total, we have eight parameters:
\begin{enumerate}[nolistsep]
\item The explosion time;
\item The binary separation, $a$;
\item $x \propto M v^7$ (Equation~(\ref{eq:x}));
\item A factor on the $r$ SiFTO template;
\item A factor on the $i$ SiFTO template;
\item The time of peak;
\item The stretch; and
\item A factor on the shock component in $U$ (see below).
\end{enumerate}

We fit this combined model to our \textit{UBVgri} light curve from LCO using a Markov Chain Monte Carlo routine based on the \texttt{emcee} package \citep{emcee}. We cannot include the \textit{Swift} data in the fit due to a lack of early UV SN~Ia templates. In addition to the photometric uncertainty, we add a 2\% systematic (0.02~mag) uncertainty in quadrature as an estimate for our calibration uncertainties. For each of our observations, we find the expected $R_\mathrm{phot}$ and $T_\mathrm{eff}$ at that time, calculate the corresponding average $L_\nu$ in each filter, and compare that to our measured $L_\nu$.

\begin{figure}
\includegraphics[width=\columnwidth]{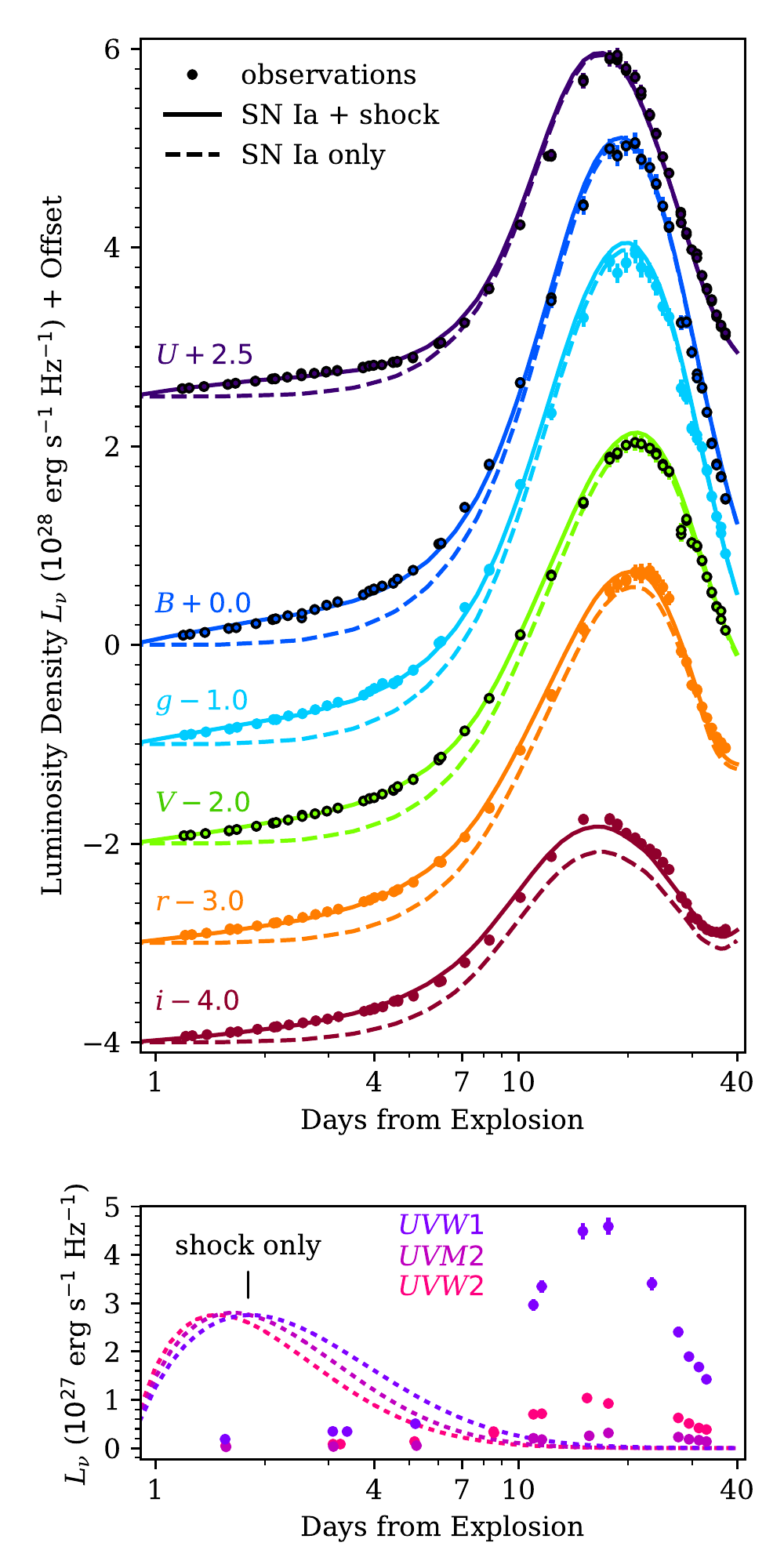}
\caption{\scriptsize Best-fit model (solid lines), consisting of an SN~Ia template \citep{sifto} plus a companion shock component from \cite{Kasen10}, to the ground-based light curve of SN~2017cbv. The dashed lines show the best-fit model with the companion shock subtracted. The model provides a good fit to the early light curve bump, but the shock model alone (dotted lines) overpredicts the early \textit{UVW1} to \textit{UVW2} luminosity.\label{fig:kasen}}
\end{figure}

One caveat to our approach is that the SiFTO template may not describe the early light curve behavior of normal SNe~Ia correctly in all filters. Given the small number of events observed 10--20 days before peak, reliable templates do not exist at these phases. However, we are encouraged by the fact that our SiFTO-based results qualitatively agree with other template fitters that we experimented with---SALT2 \citep{salt2}, MLCS2k2 \citep{mlcs2k2}, and the observed light curve of SN~2011fe \citep{Zhang16}---even at these early times.

We find that the SiFTO+\citeauthor{Kasen10} model provides an excellent fit to our ground-based data (reduced $\chi^2 = 8.6$ after including our 2\% calibration uncertainty), correctly predicting the blue bump at early times (Figure~\ref{fig:kasen}, top). Although at first glance our light curves do not look peculiar in the redder bands, emission from the shock model several weeks after explosion contributes to the observed luminosity around peak. Specifically, our best-fit model indicates that 5\% and 15\% of the $r$- and $i$-band peak luminosities, respectively, come from the shock component.

The best-fit binary separation is $56\,R_\sun$, implying a stellar radius of $\sim20\,R_\sun$ \citep[assuming Roche lobe overflow;][]{Eggleton83}. $56\,R_\sun$ is among the largest binary separations for SD SN~Ia progenitors from binary population synthesis calculations \citep{Liu15}. However, this value is quite sensitive to the early color evolution of our SN~Ia template. As these templates may not be valid 15--20 days before peak, this result should be treated with caution. Furthermore, our simplified spherical model ignores the degeneracy between binary separation ($a$) and viewing angle; a bright (large $a$) off-angle shock looks similar to a faint (small $a$) shock along the line of sight.

Given the strong dependence of $x$ on the transition velocity ($\propto v^7$), which is not observable, we cannot robustly estimate the ejecta mass. However, taking our best-fit value of $x=3.84\pm0.19$ and assuming a Chandrasekhar mass of ejecta, we find a reasonable transition velocity of $v\approx12000$~km~s$^{-1}$ (subject to uncertainty in the distance modulus). The best-fit explosion time is MJD 57821.9, about 7~hr before discovery, and the best-fit time of $B$-band peak for the SiFTO component is MJD 57840.2. The best-fit stretch from the SiFTO template is 1.04.

Despite the success of the binary companion shock model in the optical, we required a scaling factor of 0.61 on the $U$-band shock component in order to fit the data. The \textit{UVW1}, \textit{UVM2} and \textit{UVW2} emission are even further overpredicted by the shock model (Figure~\ref{fig:kasen}, bottom). We discuss the potential causes of this discrepancy in \S\ref{sec:discuss}.

\section{Early Spectra}\label{sec:spec}

\begin{figure*}[!t]
\begin{center}
\includegraphics[width=0.49\textwidth]{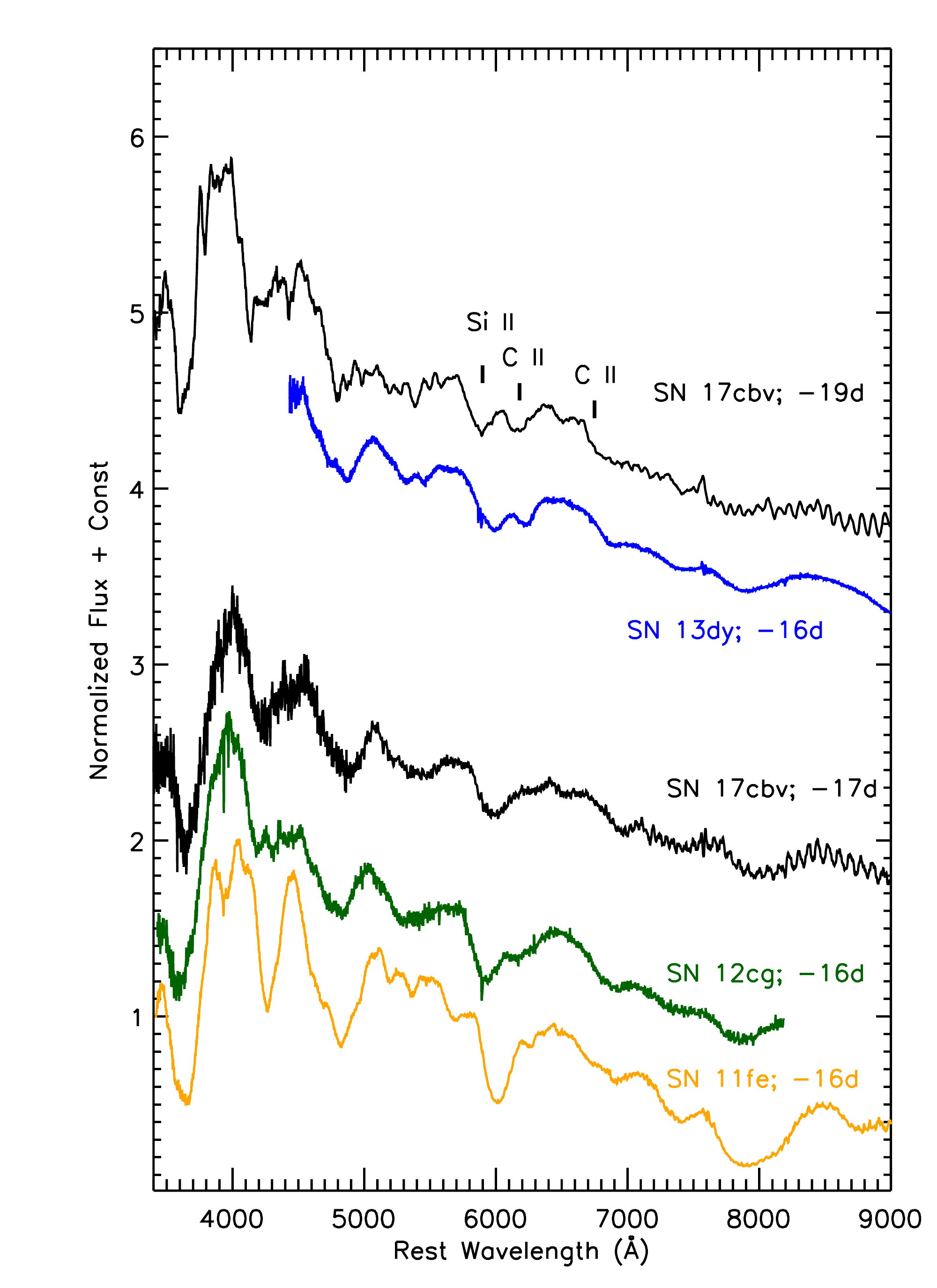}
\includegraphics[width=0.49\textwidth]{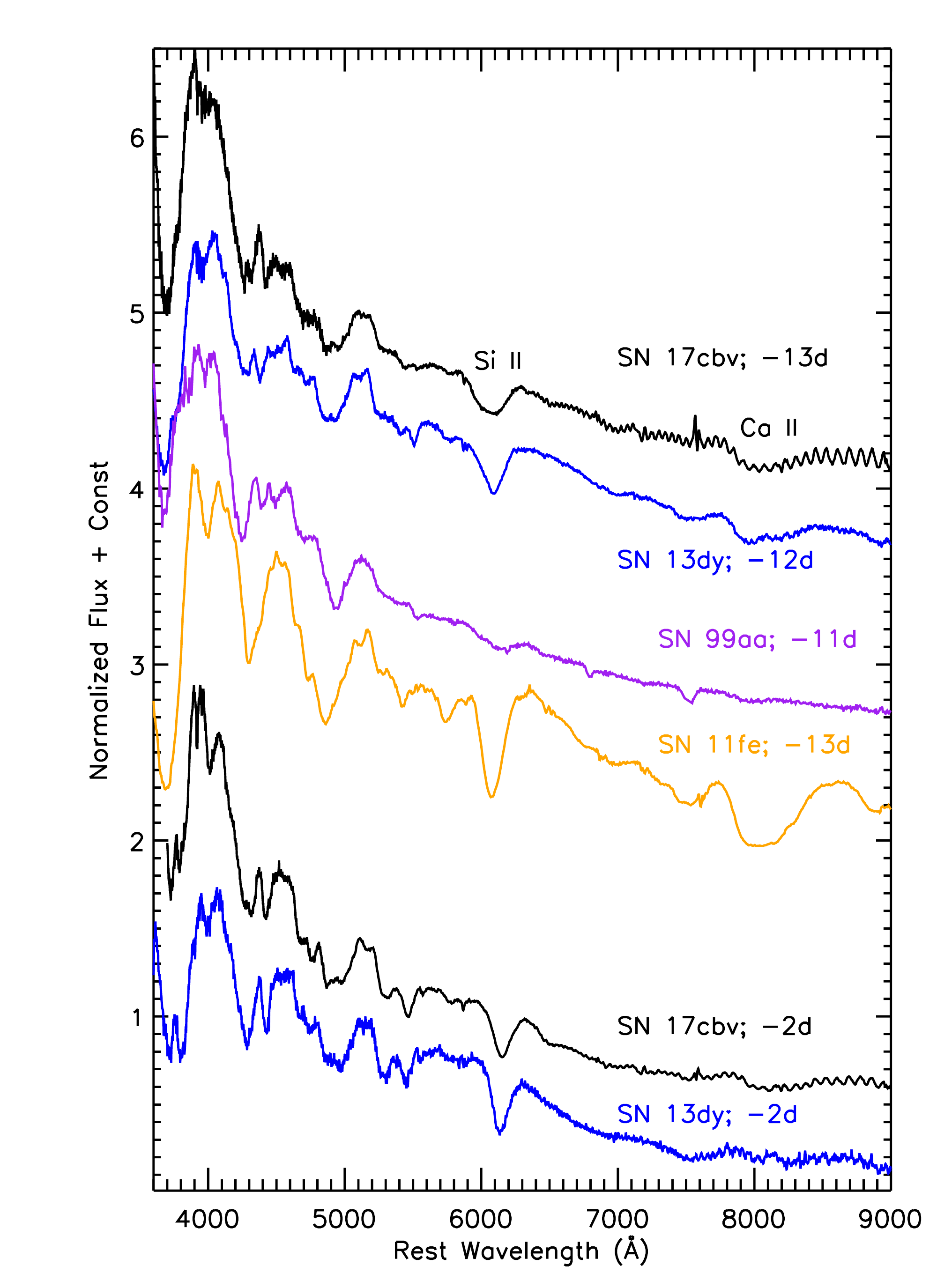}
\includegraphics[width=0.6\textwidth]{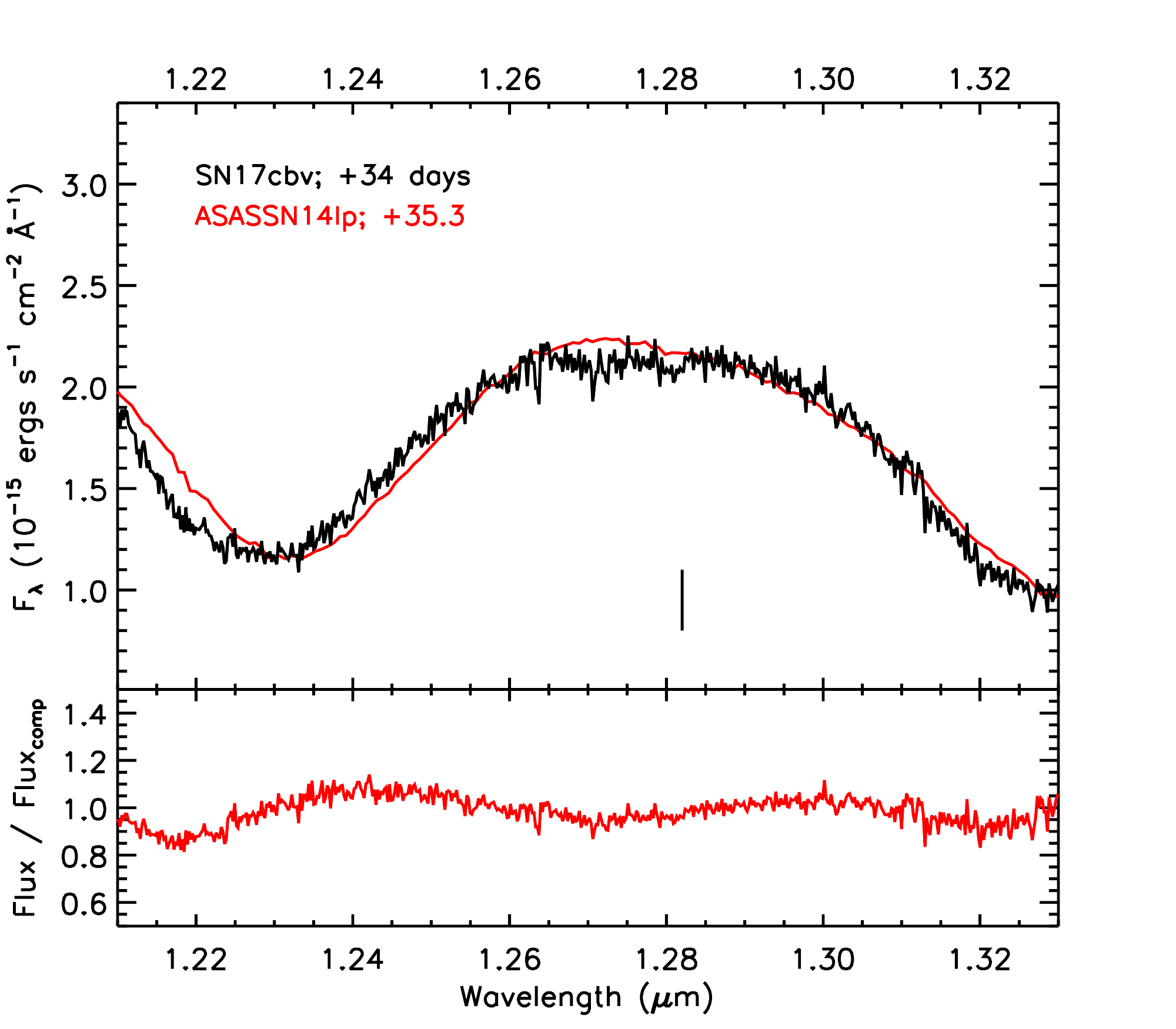}
\caption{\scriptsize Top left: the smoothed $-19$ day spectrum of SN~2017cbv shows conspicuous carbon similar to SN~2013dy, which persists at least until day $-17$.  The $-17$ day spectrum is quite similar to the $-16$ day spectrum of SN~2012cg, which also showed an early blue bump in its light curve, but not the $-16$ day spectrum of SN~2011fe. Top right: spectra of SN~2017cbv at $-13$ and $-2$ days are very similar to SN~2013dy at similar epochs.  \ion{Si}{2} and \ion{Ca}{2} are present and too strong to be SN 1999aa-like, but not as strong as in SN~2011fe. Bottom: NIR spectrum of SN~2017cbv in the region of Pa$\beta$ (marked by the vertical line) compared to a spectrum of ASASSN-14lp. The bottom panel shows the ratio of the two spectra, yielding no evidence of narrow hydrogen emission from interaction with a companion star.
}
\label{fig:spec}
\end{center}
\end{figure*}

We obtained several additional optical and near-infrared (NIR) spectra of SN~2017cbv with FLOYDS and the SpeX NIR spectrograph \citep{Rayner03} at the NASA Infrared Telescope Facility, a selection of which are presented in Figure~\ref{fig:spec}. As the event is ongoing at the time of publication, the full data set and analysis will be presented elsewhere (D.~J.\ Sand et al. 2017, in preparation). In summary, the spectrum of SN~2017cbv greatly resembles the spectrum of SN~2013dy \citep{Zheng13} during the two weeks before maximum light, with \ion{Si}{2} and \ion{Ca}{2} absorption features weaker than the prototypical Type~Ia SN~2011fe but stronger than the somewhat overluminous SN~1999aa \citep{Garavini04}. Despite the spectroscopic similarity to SN~2013dy, the light curve of SN~2017cbv declines slightly faster: $\Delta m_{15}(B)=1.06$~mag for SN~2017cbv versus 0.92~mag for SN~2013dy \citep{Pan15}.

We measure a \ion{Si}{2} $\lambda$6355 velocity of 22,800~km~s$^{-1}$ in the initial spectrum of SN~2017cbv, 19 days before maximum light. The absorption feature just redward of \ion{Si}{2} could either be a lower-velocity component (9400~km~s$^{-1}$) of the same line or \ion{C}{2} $\lambda$6580 at 19,800~km~s$^{-1}$. We prefer the latter interpretation, as (1) it correctly predicts the tentative position of \ion{C}{2} $\lambda$7234 in the same spectrum and (2) 9400~km~s$^{-1}$ would be uniquely low among very early SN~Ia spectra \citep{Silverman12}. The detection of unburned carbon can directly discriminate between the proposed SN~Ia explosion mechanisms but is rarely seen in optical spectra even at these early times \citep{Parrent11,Silverman_carbon}, except for in super-Chandrasekhar SNe~Ia \citep{Howell06}. In SN~2017cbv, this feature disappears by day $-13$, reinforcing the need for early spectroscopy to fully account for unburned carbon.

The \ion{Si}{2} absorption feature at $-13$ days clearly shows signs of two velocity components at 16,500 and 10,500~km~s$^{-1}$.  It is likely that the earlier spectra of SN~2017cbv are dominated by the high-velocity component of \ion{Si}{2}, but we cannot fully trace the transition to low-velocity \ion{Si}{2} due to our lack of FLOYDS spectra between $-13$ and $-2$ days.  We note that $-13$ days was also the approximate epoch at which SN~2012fr, another SN~Ia with a prominent high-velocity \ion{Si}{2} component, began showing low-velocity \ion{Si}{2} \citep{Childress13}.  A similar multi-component velocity structure is evident for the \ion{Ca}{2} H\&K feature in our pre-maximum spectra.

We fit the early velocity evolution of \ion{Ca}{2} H\&K and the \ion{Si}{2} high-velocity component (the only one we can clearly identify at early times) to a $t^{-0.22}$ power law, as suggested by \citet{Piro13} for finding the explosion time for SNe~Ia.  To do this, we mimicked the methodology of \citet{Piro14} and allowed the power-law dependence to vary between $t^{-0.20}$ and $t^{-0.24}$ to estimate our uncertainties. This fit implies an explosion on MJD $57821.0 \pm 0.3$, 1.1~days prior to discovery and 0.9~days before the implied explosion time from our binary shock + standard SN~Ia model presented in \S\ref{sec:lc}.

If SN~2017cbv had an SD progenitor, we might expect to see hydrogen in its late-time spectra \citep{Mattila05,Leonard07,Maeda14}. However, we do not detect Pa$\beta$ emission in an NIR spectrum taken 34 days after maximum light (Figure~\ref{fig:spec}, bottom). Following the method of \cite{Sand16}, we calculate a rough limit of $\lesssim0.1\,M_\sun$ of hydrogen by comparing to a spectrum of ASASSN-14lp at a similar phase, although this limit depends on the viewing angle.

\section{Discussion}\label{sec:discuss}
\begin{figure}
\includegraphics[width=\columnwidth]{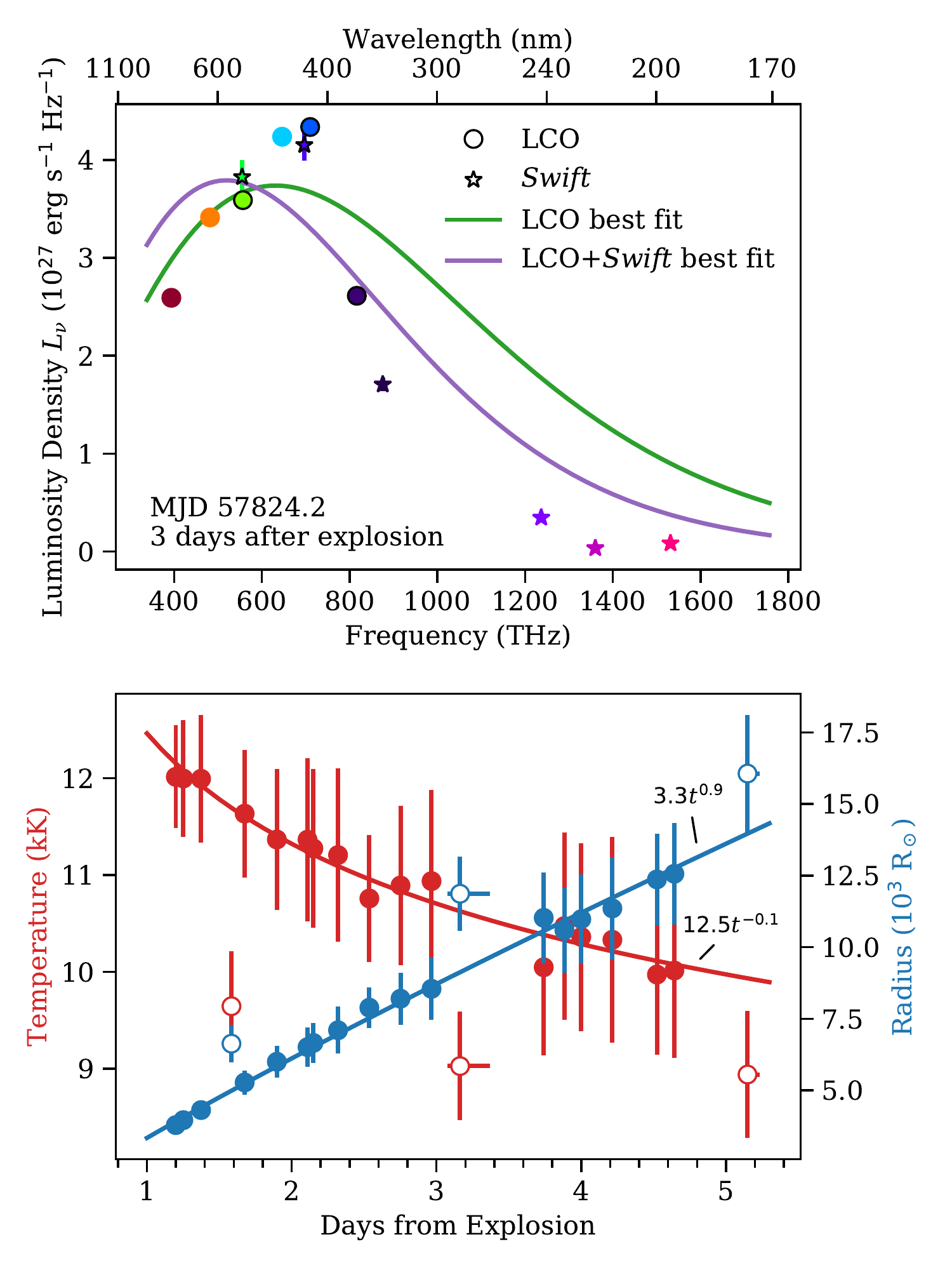}
\caption{\scriptsize Top: example SED of SN~2017cbv (three days after explosion) and two blackbody fits, with and without the \textit{Swift} observations. Including these filters pulls down the temperature, but neither blackbody spectrum is a good match to the data. Bottom: blackbody temperature and radius evolution of SN~2017cbv during the first five days after explosion, presumably when the shock component dominates the light curve. The filled points use only \textit{UBVgri} data, while the open points include \textit{Swift} photometry. Lines are best-fit power laws, excluding the open points. The temperature evolution may not be reliable, since the SED is not well-described by a blackbody, but the radius evolution is well-constrained and close to what is expected for the \cite{Kasen10} models ($t^{7/9}$).\label{fig:blackbody}}
\end{figure}

The companion-shocking models provide a good fit to our optical data, but not to our UV data. This discrepancy is not likely to be a reddening effect (unless the reddening varies very quickly with time) because the UV luminosities around peak are not unusual for SNe~Ia. However, it could stem from several simplifying assumptions in our model:

\begin{enumerate}[nolistsep]
\item \textit{Blackbody:} The analytic models assume a blackbody spectrum for the shock component. However, the observed spectral energy distribution (SED) during the bump deviates significantly from a blackbody spectrum in the UV (Figure~\ref{fig:blackbody}). A \textit{Swift} grism spectrum taken during the bump is similar to UV spectra of other SNe~Ia, showing significant absorption relative to a blackbody continuum (D.J.~Sand et al.\ 2017, in preparation). This UV suppression is likely due to line blanketing (e.g., from iron lines). Any alternative model, companion shocking or otherwise, will need to account for this deviation from a blackbody spectrum.
\item \textit{Constant Opacity:} We fixed the opacity to be that of electron scattering throughout the first 40 days of evolution, whereas in reality the opacity should change over time as the ejecta cool. Opacity and/or line blanketing that vary with time could potentially explain the discrepancy between the models and our UV data.
\item \textit{Density Profile:} The shock models we quote here assume a broken power-law density profile for the ejecta: $\rho_\mathrm{inner} \propto r^{-1}$ and $\rho_\mathrm{outer} \propto r^{-10}$. The earliest emission, which should peak in the UV bands, depends strongly on the density of the outermost ejecta layers. In particular, a steeper density profile could suppress the early luminosity.
\item \textit{Spherical Symmetry:} In order to make the problem analytically tractable, we have ignored the asymmetry that must be present in a binary system. The analytic predictions are roughly equivalent to numerical predictions for a favorable viewing angle \citep[see Figure~2 of][]{Kasen10}. If in reality we are viewing the collision off-axis, we might invoke a larger binary separation, ejecta mass, or ejecta velocity to match our observations. However, 3D numerical modeling of the ejecta--companion interaction would be needed to disentangle the early SN color diversity from the angular dependence of the shock component's color.
\end{enumerate}

If the companion-shocking scenario is correct, but the model is inaccurate in the UV, then the companion sizes found by \citet{Marion16} and \cite{Cao15} and the constraints of \citet{Brown_etal_2012_shock} would have to be reevaluated. However, if the UV overprediction is only due to line blanketing, which depends on temperature, the models may be more susceptible to failure in relatively low-temperature events. Alternatively, there could be another cause of early bumps in the UV or optical.

The observed bump could also be interpreted as a collision with CSM, rather than a collision with a companion star, as has recently been modeled by \citet{Piro16}. \cite{Kasen10} estimates that 0.01--0.1~$M_\sun$ of CSM would be needed for a significant effect, and that the material would have to be located at distances comparable to the binary separation. Stellar winds would be unlikely to produce such a configuration, but a DD system could potentially eject enough mass to large enough distances during the pre-supernova accretion phase \citep{Kasen10} or during the merger itself \citep{Levanon17}. A large mass of CSM would likely have decelerated the ejecta below the velocities we measure in \S\ref{sec:spec}, although an unshocked high-velocity component could still persist.

A third possibility is that the bump arises from a bubble of radioactive nickel that escaped most of the ejecta, allowing it to radiate light away faster than the typical diffusion timescale. \cite{Piro16} explored various nickel distributions and their effect on early SN light curves.  Shallow nickel distributions result in steeper, bluer early light curves.  The early light curves in \cite{Piro16} that included both CSM interaction and significant nickel mixing bear a qualitative resemblance to the light curve of SN~2017cbv. Likewise, \cite{Noebauer17} find an early blue bump in their sub-Chandrasekhar double-detonation model, in which the initial helium detonation on the surface produces a small amount of radioactive material. We cannot rule out these possibilities, but more detailed modeling of this event in particular would be necessary to distinguish them from the companion-shocking case.

\section{Summary}
We have presented early photometry and spectroscopy of the Type~Ia SN~2017cbv, which was discovered within $\sim$1 day of explosion. Its light curve shows a conspicuous blue excess during the first five days of observations. We find a good fit between our \textit{UBVgri} data and models of binary companion shocking from \cite{Kasen10}, but the fit overpredicts the observed UV luminosity at early times. This discrepancy might be due to several simplifying assumptions in the models. Alternatively, the excess emission could be due to interaction with CSM or the presence of radioactive nickel in the outer ejecta.

We observe no indication of ejecta interaction with hydrogen-rich material stripped from a companion star in the spectra of SN~2017cbv. However, more deep optical and near-infrared spectra out to the nebular phase are needed to confirm this finding.  Intriguingly, we do detect unburned carbon in the earliest spectra at a level rarely seen in normal SNe~Ia. A connection between an early light curve bump and the presence of unburned carbon could provide an important clue about SN~Ia progenitors, but the scarcity of events with either of these observations prevents us from drawing any conclusions now.

Our analysis demonstrates the importance of (1) discovering and announcing SNe as early as possible and (2) obtaining extremely well-sampled follow-up light curves and spectral series. The DLT40 survey and LCO follow-up network are uniquely suited to find very young SNe and follow them with sub-day cadence for long periods.  Such observations, which will become increasingly common in the coming years, greatly enhance our ability to confront theoretical models.

\acknowledgments
We thank Kyle Barbary for his help with \texttt{sncosmo} and Alex Conley and Santiago Gonz\'alez Gait\'an for providing the SiFTO templates. This work makes use of observations from the LCO network as part of the Supernova Key Project. A portion of the \textit{Swift} observations were obtained by time allocated to the Danish astronomy community via support from the Instrument Center for Danish Astrophysics (PI: Stritzinger). G.H., D.A.H., and C.M.\ are supported by the National Science Foundation (NSF) under grant No.~1313484.  D.J.S.\ and L.T.\ are supported by the NSF under grant No.~1517649. Support for S.V.\ was provided by NASA through a grant (program number HST-GO-14925.007-A) from the Space Telescope Science Institute, which is operated by the Association of Universities for Research in Astronomy, Incorporated, under NASA contract NAS5-26555. Support for I.A.\ was provided by NASA through the Einstein Fellowship Program, grant PF6-170148. E.Y.H., S.D., and M.S.\ are supported by the NSF under grant No.~AST-1613472 and by the Florida Space Grant Consortium. M.D.S.\ acknowledges support by a research grant (13261) from the Villum Fonden. D.J.S.\ is a visiting astronomer at the Infrared Telescope Facility, which is operated by the University of Hawai`i under contract NNH14CK55B with NASA. This research has made use of the NASA/IPAC Extragalactic Database, which is operated by the Jet Propulsion Laboratory, California Institute of Technology, under contract with NASA.

\bibliography{biblio}

\end{document}